%% file: main.tex
\documentclass[11pt,a4paper]{article}

\usepackage[utf8]{inputenc}
\usepackage[T1]{fontenc}
\usepackage{amsmath,amssymb,amsthm}
\usepackage{graphicx}
\usepackage{hyperref}
\usepackage{natbib}
\usepackage{geometry}
\usepackage{booktabs}
\usepackage{caption}
\usepackage{subcaption}

\title{Unambiguous Representations in Neural Networks: An Information-Theoretic Approach to Intentionality}

\author{Francesco Lässig}

\date{}

\begin{document}

\maketitle

\input{abstract}

\section{Introduction}
\input{introduction}

\section{Methods}
\input{methods}

\section{Results}
\input{results}

\section{Discussion}
\input{discussion}

\section{Conclusion}
\input{conclusion}

\bibliographystyle{plainnat}
\bibliography{references}

\input{appendix}

\end{document}

%% file: abstract.tex
\begin{abstract}
Representations pervade our daily experience, from letters representing sounds to bit strings encoding digital files. While such representations require externally defined decoders to convey meaning, conscious experience is fundamentally different: a neural state corresponding to perceiving a red square cannot alternatively encode the experience of a green triangle. This intrinsic property of consciousness suggests that conscious representations must be unambiguous in a way that conventional representations are not. We formalize this intuition using information theory, defining representational ambiguity as the conditional entropy $H(I|R)$ over possible interpretations $I$ given a representation $R$. Through experiments on neural networks trained to classify MNIST digits, we demonstrate that relational structures in network connectivity can unambiguously encode representational content. From relational structure alone, we achieve perfect (100\%) accuracy for dropout-trained networks and 38\% for standard backpropagation (chance: 10\%) in identifying output neuron class identity—despite identical task performance—demonstrating that representational ambiguity can arise orthogonally to behavioral accuracy. We further show that spatial position of input neurons—relevant to phenomenal properties like visual field location—can be decoded from network connectivity with $R^2$ up to 0.844. These results provide a quantitative method for measuring representational ambiguity in neural systems and demonstrate that neural networks can exhibit the low-ambiguity representations posited as necessary (though not sufficient) by theoretical accounts such as narrow representationalism and IIT.
\end{abstract}

%% file: introduction.tex
Conventional representations (letters, words, bit strings) are ambiguous: a bit string can be decoded as a JPEG image or an MP3 audio file depending on the decoding scheme applied. Nothing inherent in the representation determines its content; meaning depends entirely on an external decoder \citep{searle1980minds}.

Conscious brain states appear fundamentally different. If consciousness supervenes on brain states, then a neural state corresponding to perceiving a red square cannot alternatively encode perceiving a green triangle. This constraint arises because consciousness is an \emph{intrinsic} property: phenomenal experience, the ``what it is like'' quality \citep{nagel1974bat}, is determined by the neural state itself, not by external interpretation. Unlike bit strings, conscious states cannot have different contents depending on which decoder is applied.

\subsection{Narrow Representationalism and Ambiguity}

Narrow representationalism \citep{chalmers2004representational} formalizes this constraint: conscious systems instantiate intentional contents (representations of objects or features) that determine experience, and these contents are completely determined by brain states alone. Two molecularly identical brains must instantiate identical intentional contents and thus identical experiences. This means neural states \emph{unambiguously} represent their contents, in stark contrast to bit strings, where the same representation admits multiple interpretations.

We formalize ambiguity as conditional entropy:
\begin{equation}
\text{Ambiguity} = H(I|R)
\end{equation}
where $I$ denotes possible interpretations and $R$ the representation. High ambiguity corresponds to uniform distribution over interpretations (minimal information about content); low ambiguity corresponds to concentrated probability (high confidence in content).

Representations form a spectrum: bit strings are maximally ambiguous, while conscious brain states must be unambiguous (or at least significantly more so). Our hypothesis is that relational structure may be key to reducing ambiguity: content can be fixed by the web of relations a system spans internally, rather than requiring external interpretation. Structuralist approaches to consciousness \citep{lyre2022neurophenomenal, kleiner2024mathematical} suggest one avenue for how this might work: if phenomenal properties are individuated by their relational position within quality spaces, and these quality spaces are mirrored in neural relational structure, then the geometry of neural relations could directly determine what a state represents.

\subsection{Experimental Approach}

We test whether neural networks can encode information unambiguously through relational structure. Given a trained MNIST classifier \citep{lecun1998mnist}, can we determine what a neuron represents based solely on its relational position among other neurons, without knowledge of how the network was trained? If representations are truly unambiguous, relational structure alone should specify content.

We scramble neurons and ask whether their representational content can be recovered purely from relational structure: (1) Which digit class does each output neuron represent? (2) Which spatial position does each input neuron represent? These questions operationalize our theoretical framework and provide quantitative measures of representational ambiguity.

%% file: methods.tex
We conducted two experiments testing whether neural networks encode representational content unambiguously through relational structure: (1) decoding digit class from output neuron connectivity, and (2) decoding spatial position from input neuron connectivity.

\subsection{Experiment 1: Decoding Digit Class from Output Neurons}

\subsubsection{Network Architecture and Training}

We trained fully-connected networks (784-50-50-10 architecture) on MNIST digit classification under three paradigms: (1) untrained (random initialization, control), (2) standard backpropagation, and (3) backpropagation with 20\% dropout on hidden layers \citep{baldi2013understanding}. Both trained paradigms used 2 epochs to ensure learning while avoiding overfitting on this simple task; 20\% dropout represents a standard rate balancing regularization and capacity. We generated 1000 networks per paradigm using different random seeds. All networks used Adam optimizer with learning rate 0.001 and batch size 256.

\subsubsection{Decoding Task}

Can we determine which digit class an output neuron represents from connectivity alone? Let $W \in \mathbb{R}^{10 \times 50}$ denote the output layer weight matrix (rows = neurons). We randomly permute rows to create $X$ and define label $y$ as the class of the first-row neuron, ensuring neuron position conveys no information about class identity.

\subsubsection{Relational Structure Extraction}

We extract relational structure via cosine similarities, which measure directional alignment regardless of magnitude, making them robust to weight scale variations across networks. After L2-normalizing rows of $X$ to get $X_{\text{norm}}$, we compute the Gram matrix $X' = X_{\text{norm}} X_{\text{norm}}^T$, where $(X')_{i,j}$ is the cosine similarity between neurons $i$ and $j$.

\subsubsection{Decoding Methods}

\paragraph{Geometric Matching.} We construct a reference Gram matrix $G_{\text{ref}}$ by averaging the cosine similarity matrices from 5 networks with known class ordering. The logic is: if networks consistently encode the same relational structure across training runs, then a test network's Gram matrix should closely match the reference when its neurons are correctly aligned. For each of 10 test networks with unknown ordering, we search over all possible permutations of its output neurons to find the class assignment $\hat{\sigma}$ that minimizes the Frobenius distance between the permuted test Gram matrix and the reference:
\begin{equation}
\hat{\sigma} = \arg\min_{\sigma} \|G_{\text{ref}} - P_{\sigma} G_{\text{test}} P_{\sigma}^T\|_F
\end{equation}
where $P_{\sigma}$ is a permutation matrix and $\|A\|_F = \sqrt{\sum_{i,j} A_{i,j}^2}$ is the Frobenius norm. This approach tests whether relational geometry alone is sufficiently consistent across network instances to identify neuron class identity.

\paragraph{Learned Decoder.} We trained a transformer-based decoder \citep{vaswani2017attention} with self-attention layers, treating rows of $X'$ as tokens. The decoder architecture consists of 2 multi-head self-attention layers with 4 attention heads and hidden dimensionality 128, followed by a single fully-connected output layer. Crucially, we do not use positional encodings (as in the original transformer architecture) to ensure permutation invariance: the only positional information is which row corresponds to the target neuron. Training used 800 networks (8000 data points), validation used 200 held-out networks (2000 data points), ensuring the decoder learns general relational patterns. We trained for 200 epochs with batch size 64, learning rate 0.001, and cross-entropy loss. We trained 5 decoder instances with different random seeds in each setup to assess variance.

\subsubsection{Relational Structure Necessity}

To determine whether the full relational geometry is necessary or if local neighborhoods suffice, we conducted an ablation by providing the decoder with only the first row of $X'$ (the target neuron's cosine similarities with all other neurons), while masking out all pairwise similarities that do not involve the target neuron. This tests whether a neuron's class identity can be determined from its local relationships alone or requires information about the broader geometric organization.

\subsubsection{Relational Complexity and Neuron Count}

To test whether richer relational structures enable more unambiguous representations, we systematically varied the number of output neurons used to compute relational structure. From the dropout-trained 10-output-neuron networks, we selected subsets of $k$ neurons ($k = 2, 3, \ldots, 10$) and computed Gram matrices using only those neurons. For each subset size, we applied geometric matching over all $k!$ permutations of the selected neurons, scoring the position-wise accuracy of the distance-minimizing permutation exactly as in the main geometric-matching analysis, against a $1/k$ chance level. This tests whether increasing relational complexity---by adding more pairwise relations---systematically reduces representational ambiguity.

\subsubsection{Architecture Invariance}

To test whether relational structure depends on specific architectural choices, we evaluated cross-architecture transfer using the geometric matching approach. We tested three different hidden layer architectures (50-50, 25-25, and 100) in a 3$\times$3 design, constructing reference Gram matrices from networks of each architecture and decoding test networks of each architecture, yielding decoding accuracy for all nine combinations. This assesses whether representational content is encoded in an architecture-invariant manner.

\subsubsection{Dataset Discrimination}

Using the transformer-based decoder architecture described above with random output-neuron permutations, we trained a binary classifier to distinguish whether a network was trained on MNIST or Fashion-MNIST from output layer weights alone. This tests whether dataset identity is encoded in relational geometry.

\subsection{Experiment 2: Decoding Spatial Position from Input Neurons}

Can we decode which pixel position an input neuron represents from its connectivity? Using the same networks, we focus on the first-layer weight matrix $W \in \mathbb{R}^{784 \times 50}$ (columns = input neurons). After permuting columns, we decode distance from center for the target neuron, where for a pixel at grid position $(i,j)$ in the $28 \times 28$ image:
\begin{equation}
f(i,j) = \sqrt{(i-13.5)^2 + (j-13.5)^2}
\end{equation}
This measures Euclidean distance from the image center at (13.5, 13.5).

Relational structure is extracted via column-wise cosine similarities: $X' = X_{\text{norm}}^T X_{\text{norm}}$. Due to the larger neuron count (784 vs.\ 10), we use only the learned decoder (same transformer architecture, now performing regression): searching through 784! permutations is computationally intractable. Performance is evaluated using $R^2$ score. We apply the same local vs.\ global structure ablation as in Experiment 1 to assess whether the full relational geometry is necessary for spatial position decoding.

\subsection{Quantifying Ambiguity Reduction}

We compute Ambiguity Reduction Scores (ARS) to connect decoding performance with our theoretical framework:
\begin{equation}
\text{ARS} = 1 - \frac{H(I|R,C)}{H_{\max}}
\end{equation}
where $H(I|R,C)$ is conditional entropy given representation $R$ and context $C$, and $H_{\max}$ is maximum entropy. For discrete variables (Experiment 1), $H$ denotes Shannon entropy with $H_{\max} = \log_2 K$ for $K$ classes; for continuous variables (Experiment 2), $H$ denotes differential entropy with $H_{\max} = \frac{1}{2}\log_2(2\pi e \cdot \text{Var}(Y))$. While these entropy types have different mathematical properties, the normalized ratio $H/H_{\max}$ provides a comparable measure of residual uncertainty in both settings. For output neuron class decoding, we use geometric matching accuracies; for input neuron position decoding, we use learned decoder $R^2$ scores.

For classification with accuracy $A$ and $K$ classes, Fano's inequality yields:
\begin{equation}
\text{ARS} \geq 1 - \frac{h_b(1-A) + (1-A)\log_2(K-1)}{\log_2 K}
\end{equation}
where $h_b(p) = -p\log_2(p) - (1-p)\log_2(1-p)$.

For regression with $R^2$ score, we assume residuals are approximately Gaussian and standardize the target variable to unit variance ($\text{Var}(Y)=1$). Under these assumptions, the residual variance equals $1-R^2$, and the differential entropy of Gaussian residuals yields:
\begin{equation}
\text{ARS} \geq \frac{\log_2[1/(1-R^2)]}{\log_2(2\pi e)} \approx \frac{\log_2[1/(1-R^2)]}{4.094}
\end{equation}

%% file: results.tex
\subsection{Experiment 1: Output Neuron Class Decoding}

\subsubsection{Learned Decoder Performance}

\begin{figure}[htbp]
    \centering
    \includegraphics[width=0.7\textwidth]{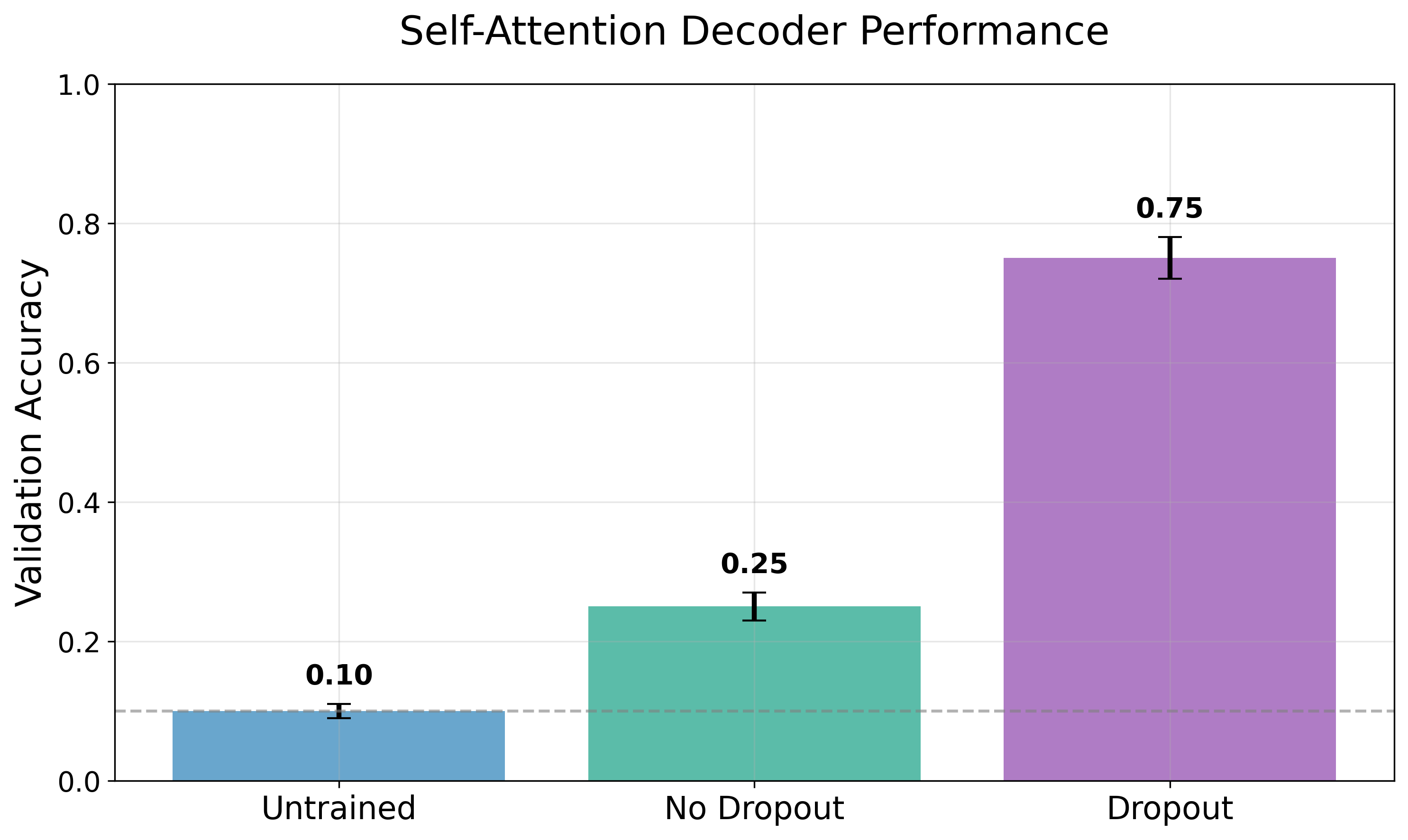}
    \caption{Decoder validation accuracy for identifying output neuron class identity across training paradigms. The decoder achieves approximately 25\% accuracy when decoding from standard backpropagation networks and 75\% when decoding from dropout networks, compared to 10\% chance level (observed when decoding from untrained networks). Error bars represent standard deviation across 5 random seeds.}
    \label{fig:decoder-accuracy}
\end{figure}

Output neuron class identity can be decoded from relational structure well above chance (Figure~\ref{fig:decoder-accuracy}).

\begin{figure}[htbp]
    \centering
    \includegraphics[width=0.65\textwidth]{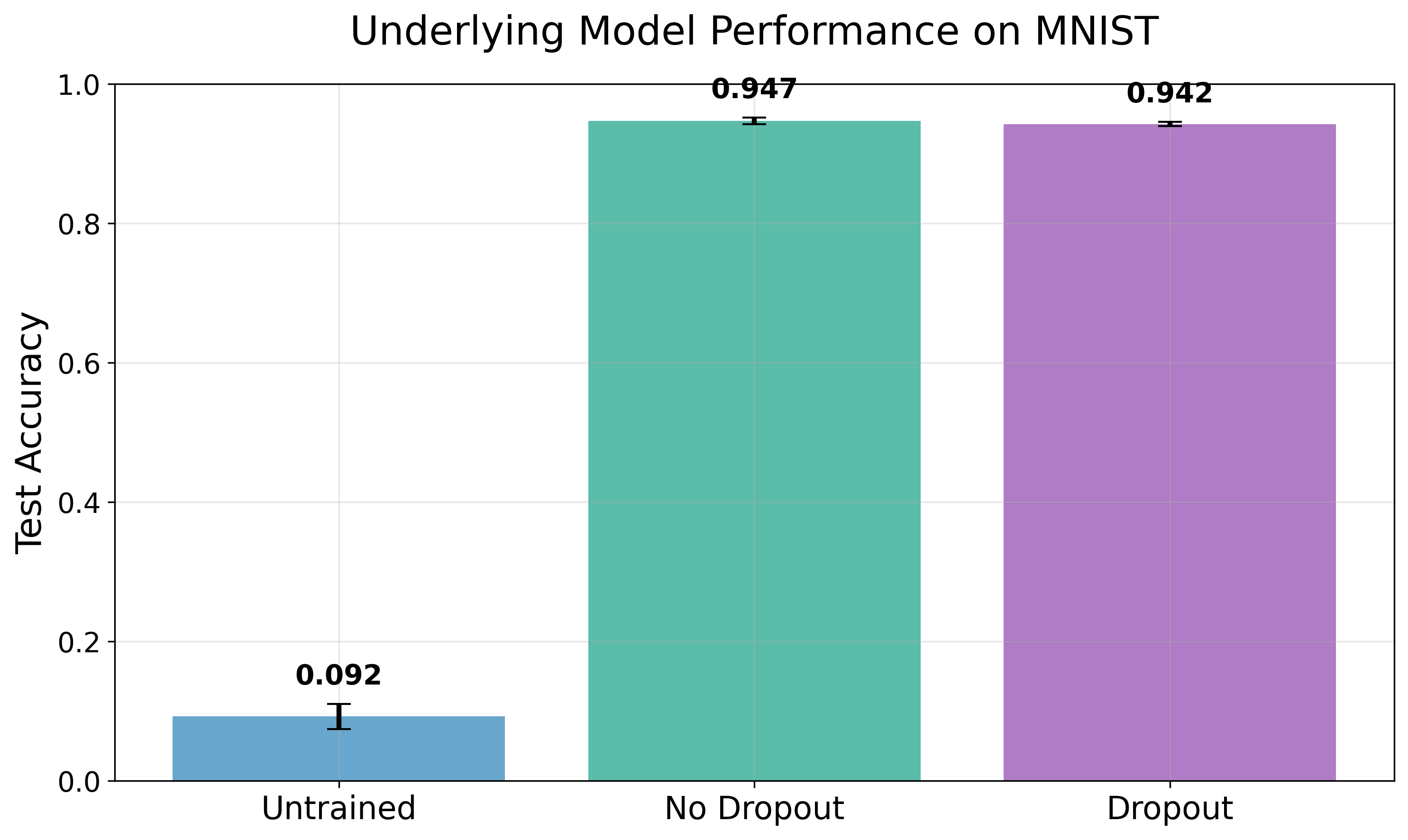}
    \caption{Validation accuracies of underlying MNIST models. Despite similar classification performance, dropout and standard training produce dramatically different levels of representational ambiguity.}
    \label{fig:mnist-accuracy}
\end{figure}

When decoding from untrained networks, the decoder achieves 10\% accuracy (chance level), validating our experimental design. For standard backpropagation networks, decoder accuracy reaches 25\% (above chance), while for dropout networks it achieves 75\%, a threefold improvement. Notably, the base networks' MNIST classification performance is nearly identical regardless of dropout (Figure~\ref{fig:mnist-accuracy}), showing that representational ambiguity is orthogonal to task performance.

\subsubsection{Geometric Matching Results}

\begin{figure}[htbp]
    \centering
    \includegraphics[width=0.65\textwidth]{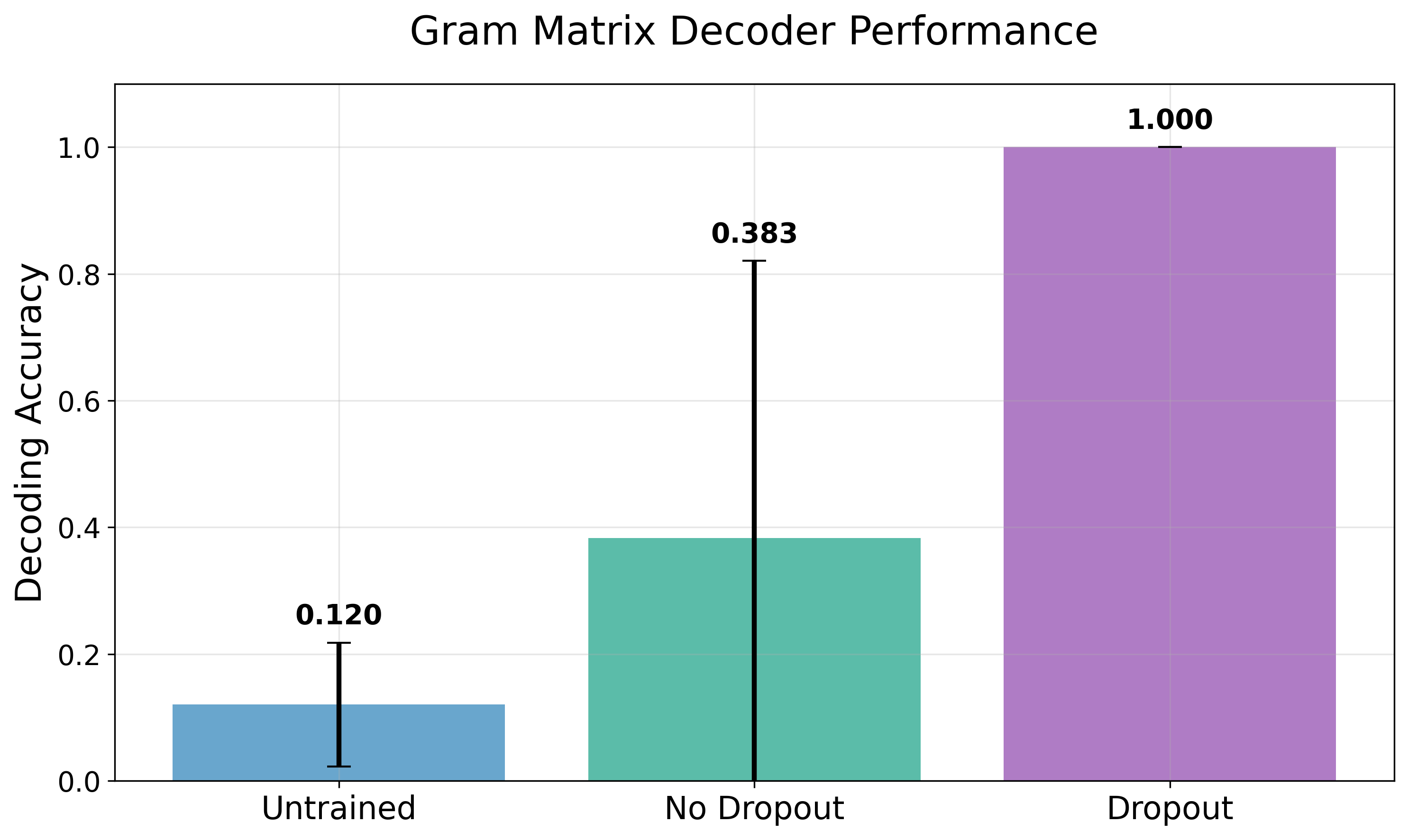}
    \caption{Geometric matching accuracies for decoding output neuron class identity across validation networks. Error bars reflect variance across network instances. Decoding from dropout networks achieves perfect 100\% accuracy with zero variance, indicating completely unambiguous relational encoding of class identity. Larger error bars for standard backpropagation reflect geometric ambiguity: multiple permutations achieve near-identical distances to the reference.}
    \label{fig:gram-accuracy}
\end{figure}

Geometric matching (directly comparing relational geometries without training) achieves even higher accuracies (Figure~\ref{fig:gram-accuracy}).

Geometric matching achieves perfect 100\% decoding accuracy for dropout networks with zero variance, demonstrating that their relational geometries unambiguously specify class identity. This result shows that the relational structure alone—without any information about neuron ordering or training procedure—completely determines what each neuron represents.

\begin{figure}[htbp]
    \centering
    \begin{subfigure}[b]{0.7\textwidth}
        \includegraphics[width=\textwidth]{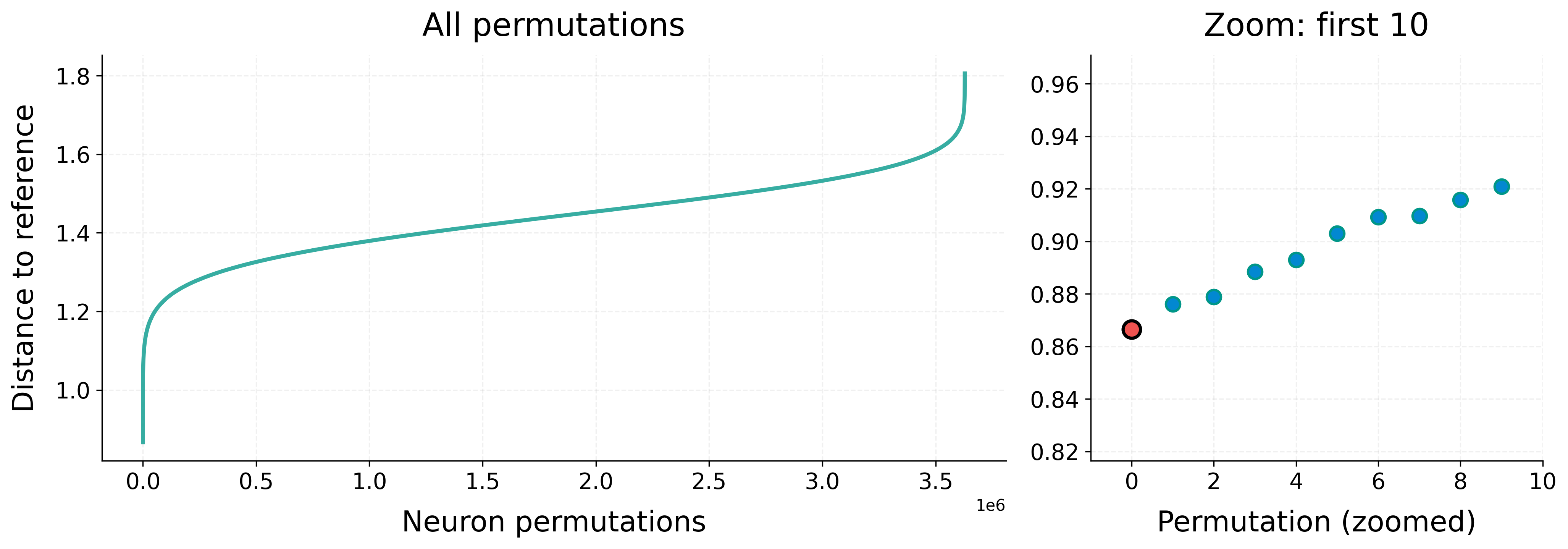}
        \caption{Standard backpropagation}
    \end{subfigure}

    \vspace{1em}

    \begin{subfigure}[b]{0.7\textwidth}
        \includegraphics[width=\textwidth]{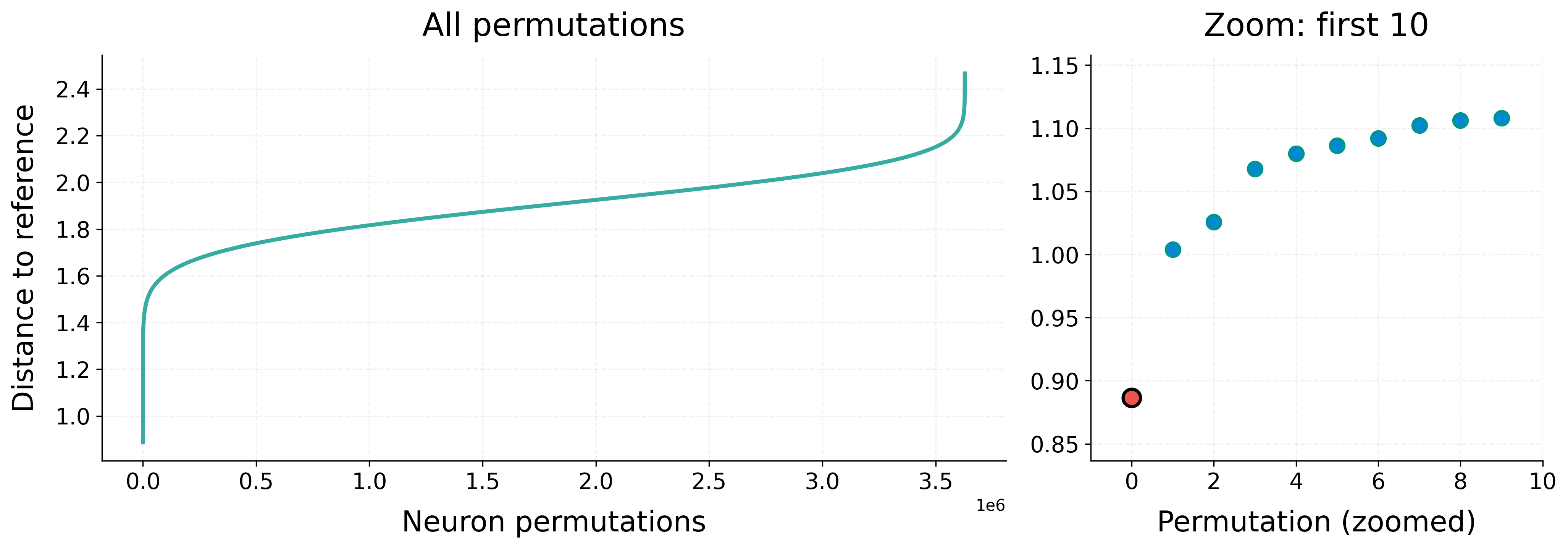}
        \caption{Dropout training}
    \end{subfigure}
    \caption{Frobenius distances between test and reference Gram matrices for all permutations of test network output neurons. Red dots indicate the true permutation. For standard backpropagation (a), even when the true permutation achieves the lowest distance, it does so by only a small margin. Dropout training (b) creates clear separation.}
    \label{fig:perm-distances}
\end{figure}

Figure~\ref{fig:perm-distances} reveals why geometric matching succeeds or fails. For two test networks (one trained with dropout, one trained with standard backpropagation), we compute the Frobenius distance between the reference Gram matrix and the test network's Gram matrix under all possible neuron permutations (10! = 3,628,800 permutations). If relational structure unambiguously encodes class identity, the true permutation (red dot) should yield a substantially lower distance than all incorrect permutations. For dropout networks, this is exactly what we observe: the true permutation is clearly separated from all alternatives, creating an unambiguous geometric signature. In contrast, for standard backpropagation networks, the true permutation shows only a small margin over many incorrect alternatives, making reliable identification difficult. This geometric clarity in dropout networks explains why both our learned decoder (75\% accuracy) and geometric matching (100\% accuracy) achieve their best performance on these networks.

\subsubsection{Relational Structure Necessity}

\begin{figure}[htbp]
    \centering
    \includegraphics[width=0.65\textwidth]{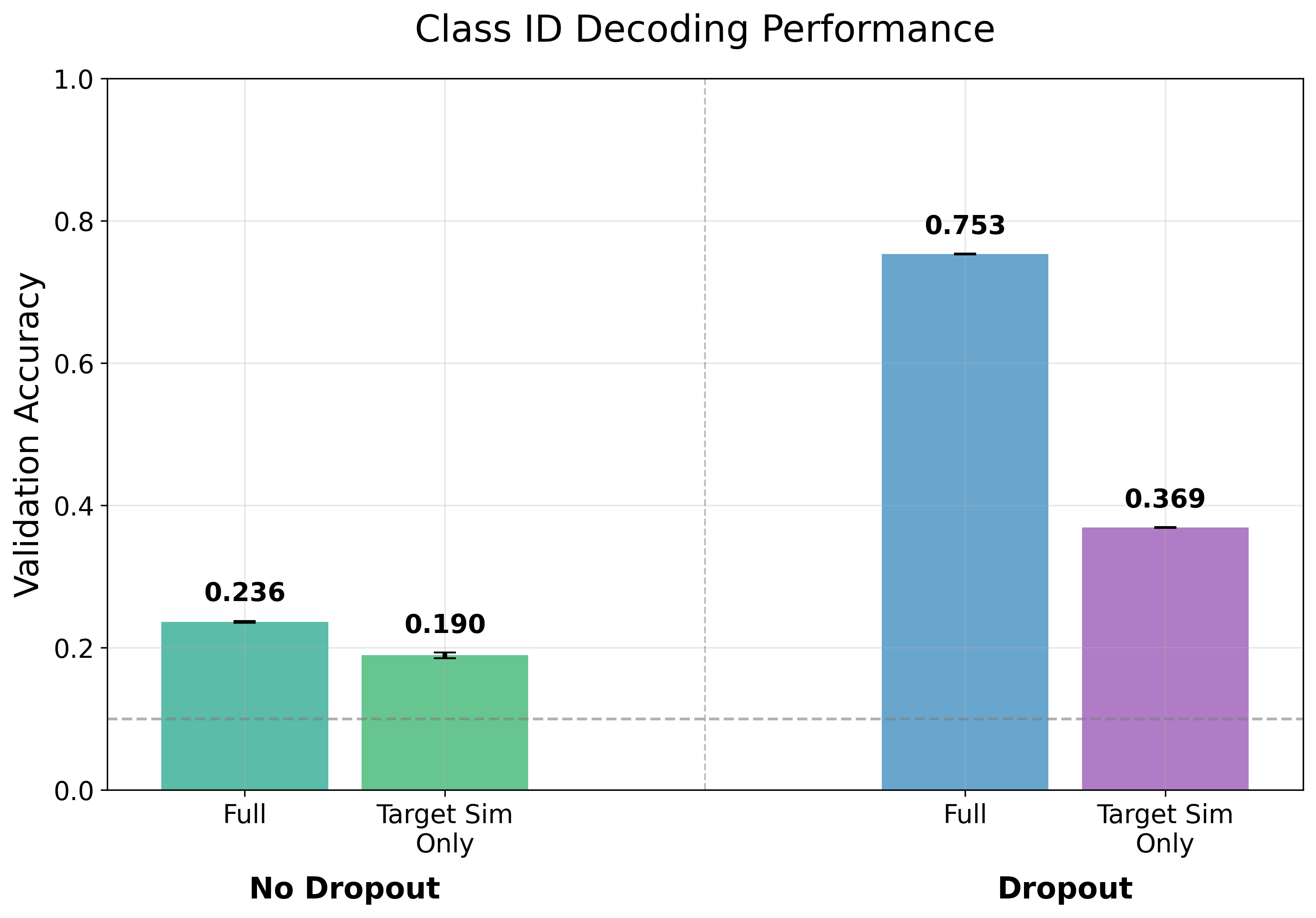}
    \caption{Decoder accuracy when provided only the target neuron's local neighborhood versus full relational structure. The complete geometry is essential for accurate decoding.}
    \label{fig:target-only-output}
\end{figure}

To determine whether the entire relational geometry is necessary or if local neighborhoods suffice, we provided the decoder with only the target neuron's similarity vector to all other neurons, masking out pairwise similarities between non-target neurons. Figure~\ref{fig:target-only-output} shows that decoding accuracy drops substantially when this broader geometric context is removed. This demonstrates that a neuron's representational content cannot be determined from its local relationships alone—the decoder requires information about how the entire output population is organized relative to each other.

\subsubsection{Relational Complexity Scaling}

\begin{figure}[htbp]
    \centering
    \includegraphics[width=0.7\textwidth]{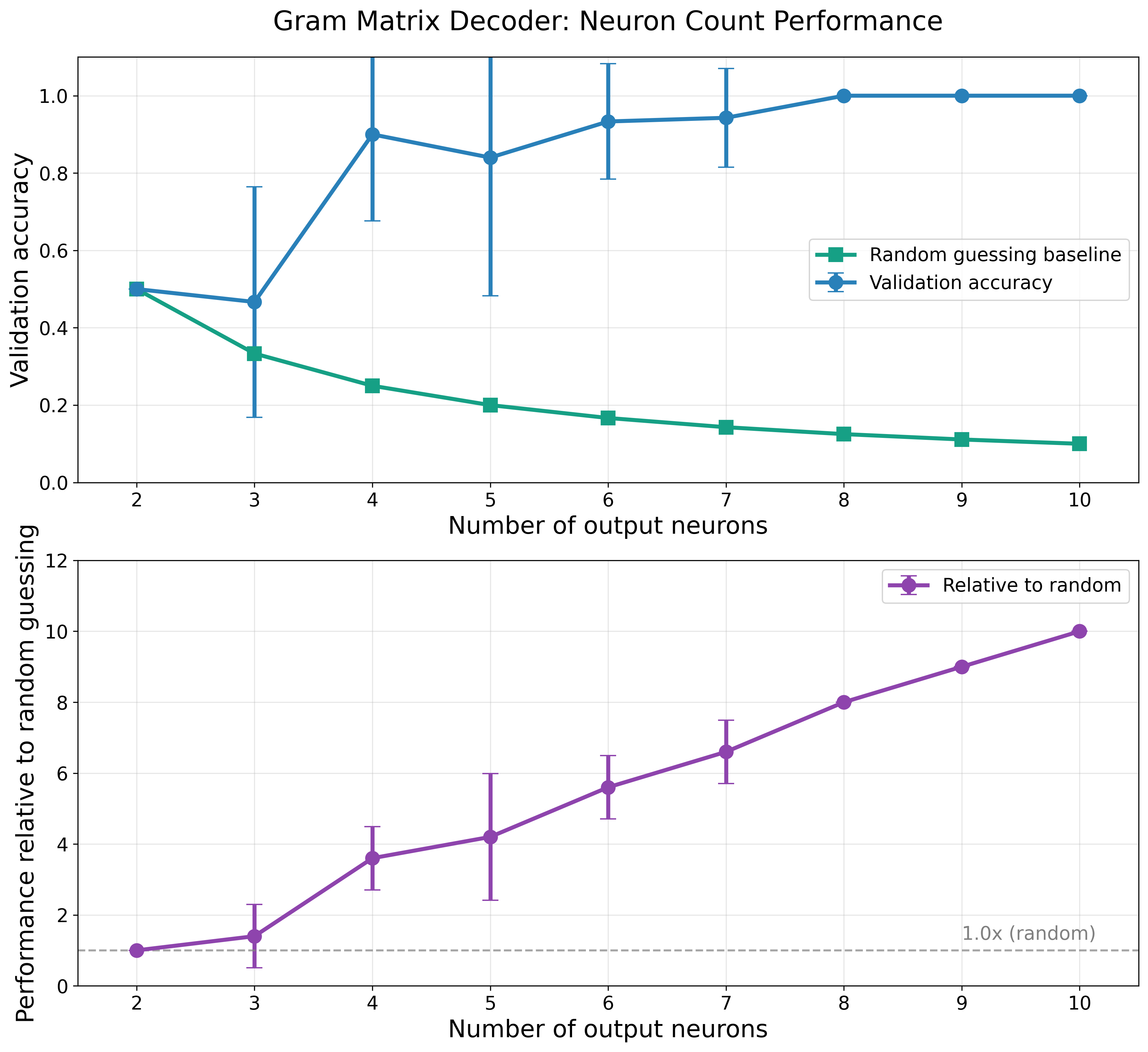}
    \caption{Geometric matching performance as a function of relational structure complexity, using the dropout networks and the same accuracy metric as Figure~\ref{fig:gram-accuracy}. The top panel shows validation accuracy (blue) versus random guessing baseline (teal) for 2-10 output neurons. Error bars represent standard deviation across 5 random seeds. The bottom panel shows performance relative to random chance (purple). The 2-neuron case achieves only random-level performance (1.0x), since asymmetric relational structures cannot exist between two points. Performance increases systematically with neuron count, reaching 100\% accuracy and 10x better than random for the 10-neuron model.}
    \label{fig:ablation-neuron-count}
\end{figure}

Beyond establishing that full relational geometry is necessary, we tested whether richer relational structures systematically reduce ambiguity. Using the dropout networks from the main geometric-matching result, we varied the number of output neurons used to compute relational structure from 2 to 10, applying geometric matching to Gram matrices of varying complexity and scoring the same way as in \S3.1.2 (Figure~\ref{fig:ablation-neuron-count}).

The 2-neuron case performs exactly at chance level (50\%, or 1.0x), confirming that asymmetric relational structures are essential for decoding. Performance relative to chance then increases monotonically with neuron count, from 1.0x (2 neurons) to 10x (10 neurons), where absolute accuracy reaches 100\%---matching the unambiguous encoding observed for the full 10-neuron dropout networks in \S3.1.2. This systematic improvement demonstrates that richer relational geometries provide more constraints on possible interpretations, thereby reducing representational ambiguity.

\subsubsection{Architecture Invariance}

\begin{figure}[htbp]
    \centering
    \includegraphics[width=0.7\textwidth]{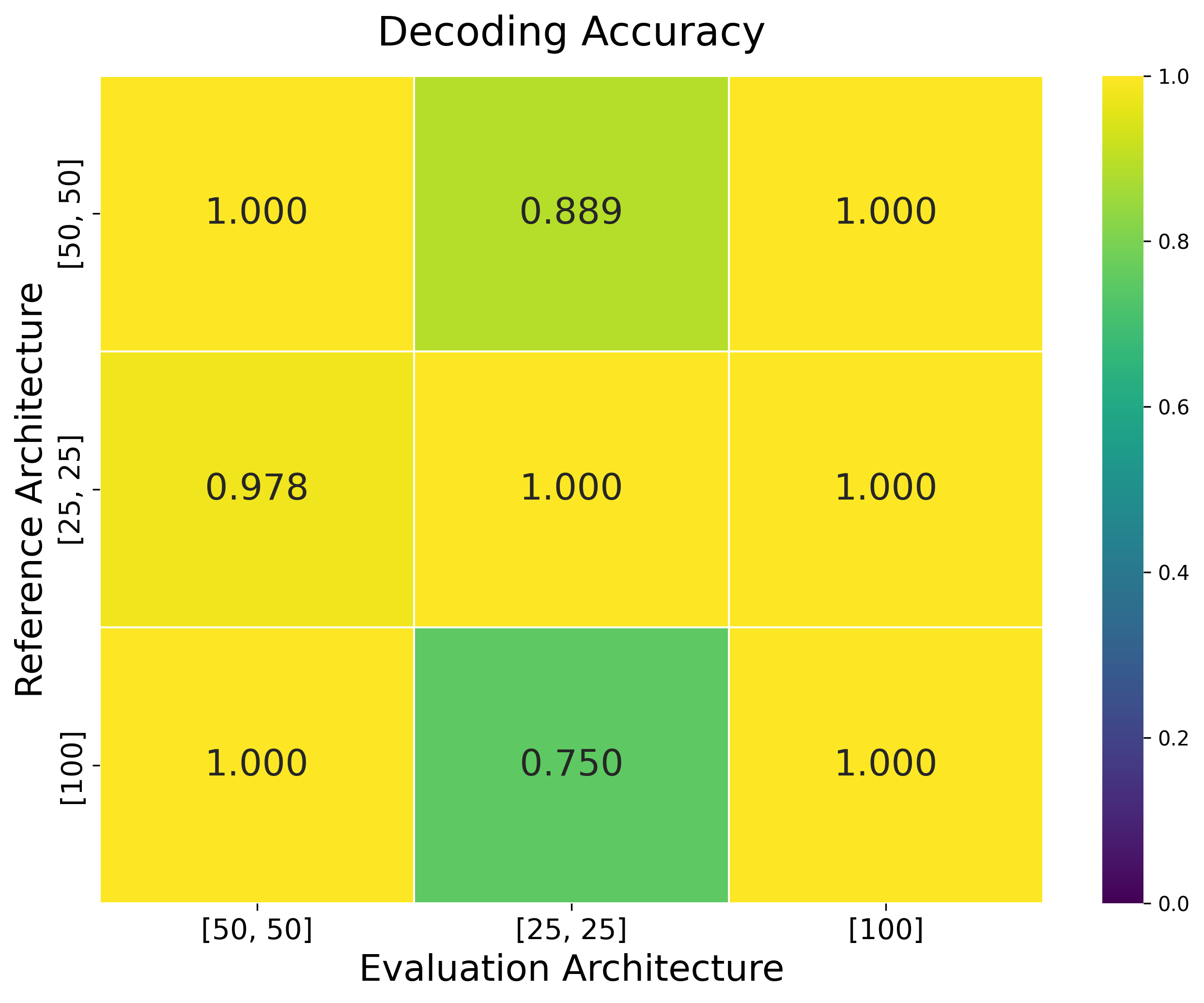}
    \caption{Cross-architecture transfer accuracy. Strong performance across different architectures demonstrates that relational structure is largely architecture-invariant.}
    \label{fig:cross-arch}
\end{figure}

A key question is whether relational structure depends on specific architectural choices or represents a more fundamental property. We evaluated cross-architecture transfer using geometric matching in a 3$\times$3 design, testing three different hidden layer architectures (50-50, 25-25, and 100) as both reference and test networks. Figure~\ref{fig:cross-arch} shows strong decoding performance across all architectural combinations, demonstrating that the relational geometry encoding class identity is largely architecture-invariant.

\subsubsection{Dataset Discrimination}

Dataset identity (MNIST vs.\ Fashion-MNIST) can be classified from relational structure with near-perfect accuracy for dropout (99.8\% $\pm$ 0.1\%) and clearly above chance for standard training (84.3\% $\pm$ 0.8\%). This shows that dataset identity is encoded in relational geometry.

\subsection{Experiment 2: Input Neuron Spatial Position Decoding}

\begin{figure}[htbp]
    \centering
    \includegraphics[width=0.7\textwidth]{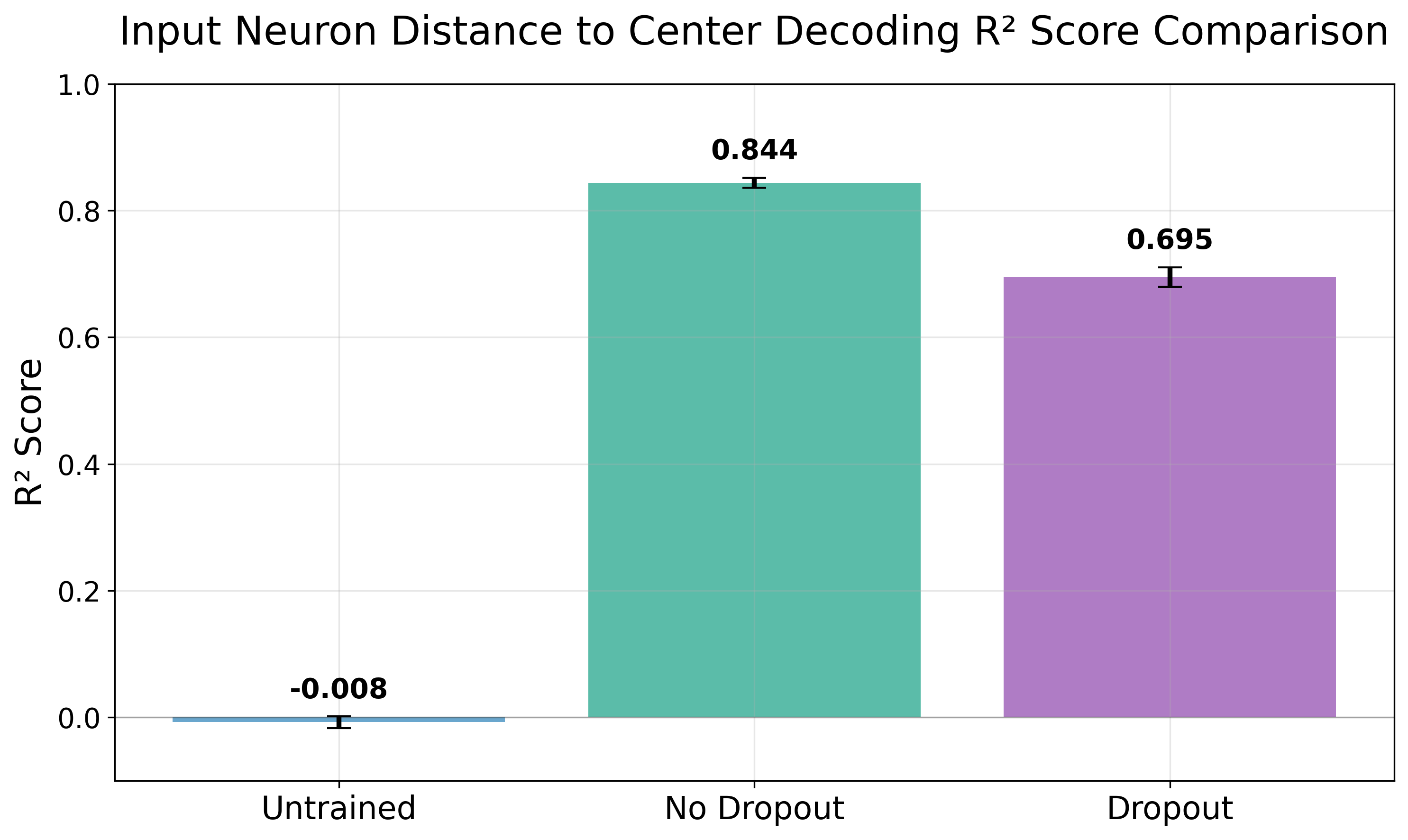}
    \caption{Decoder $R^2$ score for predicting input neuron distance from center. Unlike output neurons, decoding from standard backpropagation networks yields higher accuracy than decoding from dropout networks for this task.}
    \label{fig:input-distance}
\end{figure}

While digit class identity is abstract, spatial position provides a more direct connection to phenomenal properties like visual field location. We asked whether input neurons encode their pixel positions through relational structure. Figure~\ref{fig:input-distance} shows that they do: the decoder achieves $R^2 = 0.844$ when decoding from standard backpropagation networks and $R^2 = 0.695$ when decoding from dropout networks, both substantially above the untrained baseline ($R^2 \approx 0$).

\begin{figure}[htbp]
    \centering
    \includegraphics[width=0.65\textwidth]{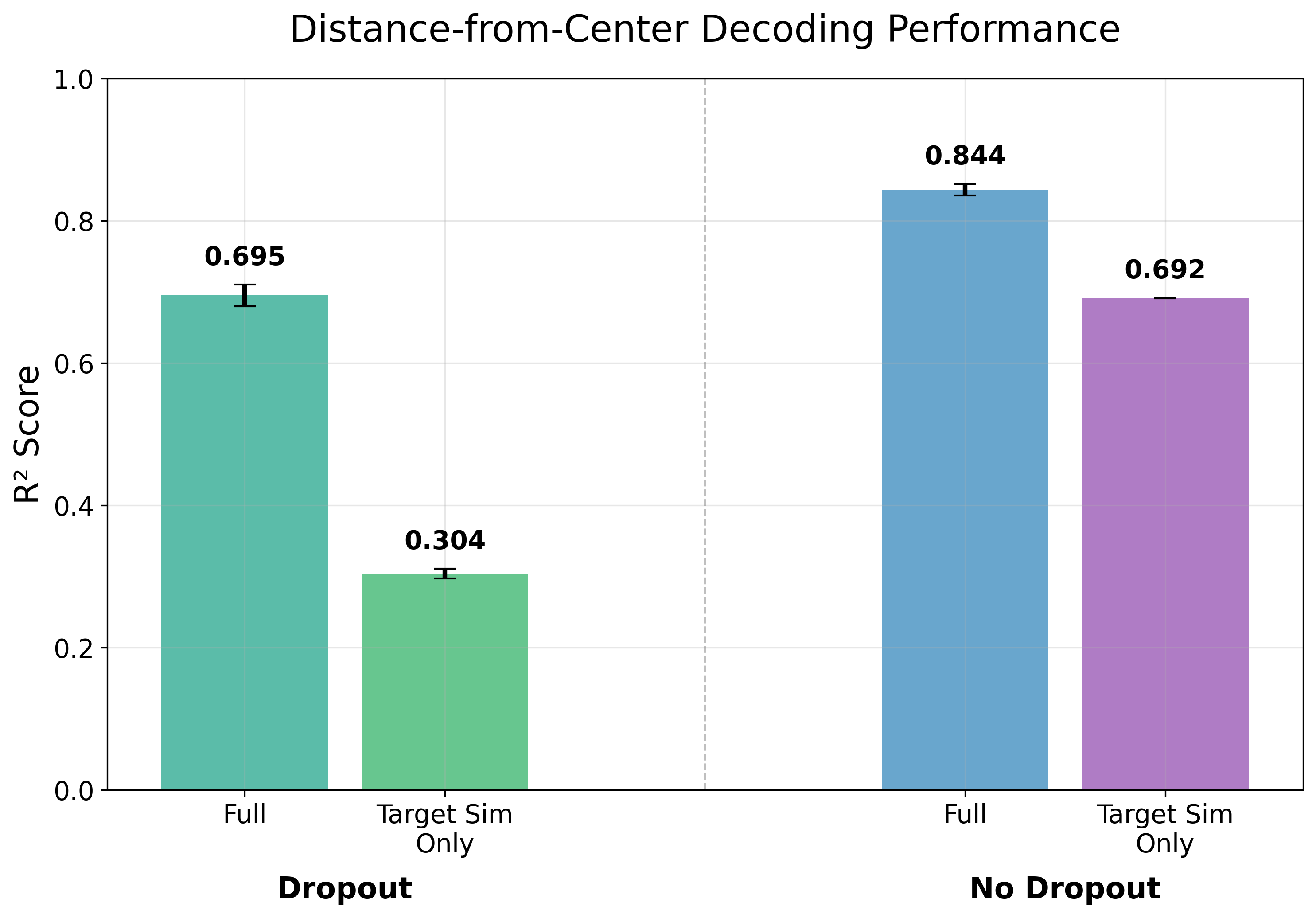}
    \caption{Decoder performance for input neurons when using only local neighborhood versus full relational structure.}
    \label{fig:target-only-input}
\end{figure}

Interestingly, the pattern reverses compared to output neurons: standard backpropagation networks yield higher spatial decoding accuracy than dropout networks. The precise mechanism underlying this reversal remains unclear, though it may reflect different optimization dynamics affecting input versus output layers under dropout training. Nevertheless, as in Experiment 1, providing the decoder with the full relational structure substantially outperforms using only local neighborhoods (Figure~\ref{fig:target-only-input}), confirming that broader geometric context enhances decoding across different representational domains.

\subsection{Ambiguity Reduction Scores}

Table~\ref{tab:ars-scores} summarizes the Ambiguity Reduction Scores (ARS) computed from our experimental results.

\begin{table}[htbp]
    \centering
    \begin{tabular}{llcc}
        \toprule
        Experiment & Training & Performance & ARS (lower bound) \\
        \midrule
        Output (class) & Dropout & Acc = 1.000 & 1.000 \\
        Output (class) & Standard & Acc = 0.383 & 0.122 \\
        Output (class) & Untrained & Acc = 0.120 & 0.001 \\
        \midrule
        Input (position) & Dropout & $R^2 = 0.695$ & 0.419 \\
        Input (position) & Standard & $R^2 = 0.844$ & 0.654 \\
        Input (position) & Untrained & $R^2 = -0.008$ & 0.000 \\
        \bottomrule
    \end{tabular}
    \caption{Ambiguity Reduction Scores. An ARS of 1.0 indicates complete disambiguation, zero conditional entropy. Dropout networks achieve perfect disambiguation for output neuron class identity.}
    \label{tab:ars-scores}
\end{table}

The perfect ARS of 1.0 for dropout output neurons indicates $H(I|R,C) = 0$: relational structure completely determines representational content within the experimental context. Even for the more challenging spatial position task, we achieve ARS values of 0.419–0.654, demonstrating substantial ambiguity reduction.

It is important to note that these are lower bounds based on Fano's inequality and Gaussian assumptions, which may be conservative. The true ambiguity reduction could be higher. For instance, an ARS of 0.654 (input neurons, standard backpropagation) indicates that we have reduced conditional entropy by at least 65.4\% relative to maximum entropy, but the actual reduction may be greater since the bounds can be loose. Nevertheless, even these conservative estimates demonstrate substantial disambiguation, providing quantitative support for the theoretical claim that neural representations can approach the unambiguous encoding required by theories of consciousness.

%% file: discussion.tex
\subsection{Implications for Consciousness Research}

Our experimental results demonstrate that neural networks can encode information through relational structure in a way that allows unambiguous decoding of representational content. For dropout-trained networks, we achieved perfect (100\%) accuracy in identifying output neuron class identity from connectivity patterns alone. This operationalizes the theoretical claim that conscious representations must be unambiguous: the neural state itself specifies what it represents.

A potential objection is that our decoding methods constitute external decoders. However, unlike arbitrary conventions (e.g., JPEG), our decoders \emph{discover} structure rather than impose it. Both geometric matching and learned decoders infer consistent relational patterns from network instances and generalize across architectures. This demonstrates that the decoder is not arbitrary but extracts structure implicitly present in the networks—the content is structurally determined, not conventionally assigned.

We should clarify what "unambiguous" means in this context, as it refers to informational determinacy rather than phenomenal clarity. A neural state has unambiguous representational content when that content is fixed and determinate. Consider peripheral vision: the phenomenal character itself lacks detail, yet the neural state has determinate content representing "indeterminate peripheral presence." Our framework addresses semantic determinacy (what the representation is about) rather than phenomenal determinacy (whether the experience feels clear).

How does this connect to consciousness? According to representationalist theories \citep{lycan2019representational,pennartz2018consciousness}, conscious experience is determined by intentional contents: what the brain represents. Our work adds a constraint on the \emph{manner} in which these contents must be represented: they must be represented unambiguously. This constraint resonates with Chalmers' notion of \emph{impure representationalism} \citep{chalmers2004representational}, where phenomenal consciousness requires not merely representing a content, but representing it in a particular manner.

Importantly, while we have framed this work within narrow representationalism, the core constraint emerges more generally from consciousness being intrinsic. The same intuition can be expressed using IIT terminology \citep{tononi2016integrated}: consciousness is both \emph{informative} and \emph{intrinsic}, implying that the neural substrate must unambiguously specify conscious contents. Our spatial position decoding results connect to IIT-based analyses of how spatial experience is structured \citep{haun2019space}. We acknowledge that narrow representationalism is controversial, so our framework should be understood conditionally: \emph{if} conscious experience is determined by brain-internal states alone, \emph{then} those states must encode content unambiguously.

\subsection{Training Paradigm Effects}

A striking finding is that dropout training produces dramatically more unambiguous representations than standard backpropagation for output neurons (100\% vs.\ 38\% geometric matching accuracy), despite nearly identical task performance. This dissociation suggests that representational ambiguity is largely orthogonal to classification accuracy: a network can perform well while encoding information ambiguously.

Intuitively, dropout encourages distributed representations by preventing reliance on single-neuron pathways \citep{baldi2013understanding}. This plausibly forces output neurons representing similar classes to have similar input weight patterns, as they must share overlapping features. The result is a more distinctive relational geometry where each class occupies a unique, well-separated position.

We do not claim that dropout-trained networks are more conscious than standard networks, or that consciousness requires dropout specifically. Rather, dropout serves as an existence proof that training regimes can produce varying levels of representational ambiguity while maintaining task performance. The specific mechanism is less important than the principle: ambiguity is a dimension of neural representation that can be optimized independently of behavioral performance, at least within the performance regime we tested. Whether this independence holds more generally—particularly at the limits of task complexity or performance—remains an open question.

Interestingly, the pattern reverses for input neurons: standard backpropagation yields higher spatial position decoding accuracy than dropout ($R^2 = 0.844$ vs.\ $R^2 = 0.695$). We do not speculate extensively on the mechanism, but note that different layers may be subject to different optimization dynamics. The key point is that training paradigm substantially affects representational ambiguity in ways that are not captured by task performance metrics.

\subsection{Broader Connections}

Our methodology relates to representational similarity analysis (RSA) in computational neuroscience \citep{kriegeskorte2008representational,yamins2016using} and recent work on representation alignment across networks \citep{huh2024platonic}. We extend RSA by comparing similarity structures across network instances rather than stimuli, and by using this comparison to decode specific content rather than merely assess alignment.

\subsection{Limitations and Future Directions}

Our experiments have several important limitations. First, we focus exclusively on feedforward networks trained with supervised learning on MNIST, a simple, well-structured task. Whether similar unambiguous encoding emerges in more complex scenarios remains open: unsupervised or self-supervised learning, naturalistic datasets with richer statistical structure, recurrent networks with temporal dynamics, or deeper hierarchical architectures.

Second, we measure $H(I|R,C)$ rather than $H(I|R)$, where $C$ represents context (neural network domain, dataset, architecture, decoding target). Does this fundamentally limit our conclusions? Several lines of evidence suggest the limitations are practical rather than fundamental. Dataset identity itself can be inferred from relational structure (99.8\% accuracy), suggesting the $I$/$C$ distinction is somewhat artificial. Cross-architecture transfer suggests that architecture is not a critical component of context. A universal decoder trained on diverse, multi-modal neural networks could potentially approach true $H(I|R)$ by learning to infer context from structure itself.

Third, our experiments focused on structural connectivity (synaptic weights) rather than functional connectivity (activity correlations). While functional connectivity may be more relevant for consciousness due to its dynamic, state-dependent nature \citep{pennartz2009identification}, structural connectivity provides a stable proxy in these simple feedforward networks where structural and functional patterns are tightly coupled. For trained networks in a stable state, structural connectivity captures the learned representational geometry. Nevertheless, future work should analyze functional connectivity patterns during actual stimulus processing, particularly in biological neural recordings.

Fourth, our analysis examines single layers in isolation. A complete understanding would require examining how relational structures compose across layers and whether hierarchy introduces or reduces ambiguity. Our relational complexity scaling results reveal an important principle: more complex relational structures enable more unambiguous representations. Performance relative to random chance increases systematically from 1.0x (2 neurons) to 10x (10 neurons), demonstrating that richer relational geometries provide more constraints on possible interpretations, thereby reducing representational ambiguity.

Finally, the ultimate test would be applying this framework to biological neural data. Can we measure representational ambiguity in visual cortex recordings, and does it correlate with conscious perception? Do neural populations representing consciously perceived stimuli exhibit lower ambiguity than those processing unconscious information? These questions could provide empirical bridges between computational principles and phenomenal experience.

%% file: conclusion.tex
We have established three main contributions:

\textbf{Theoretical framework.} We formalized the constraint that conscious representations must be unambiguous, emerging from consciousness being an intrinsic property. We defined ambiguity as conditional entropy $H(I|R)$ and proposed that relational structure can reduce this entropy by constraining possible interpretations. This framework does not require commitment to representationalism specifically; the same constraint follows from the intrinsicality and informativeness axioms of theories like IIT.

\textbf{Experimental operationalization.} We demonstrated that representational ambiguity can be measured empirically through decoding tasks. Networks trained with dropout achieve perfect (100\%) accuracy in class identity decoding and high accuracy (up to $R^2 = 0.844$) in spatial position decoding, corresponding to Ambiguity Reduction Scores approaching 1.0. The methodology extends naturally to other neural systems and representational domains.

\textbf{Practical insights.} Training paradigm dramatically affects representational ambiguity largely independently of task performance, at least within our tested regime. Dataset identity can be inferred from relational structure, and decoders generalize across different network architectures, suggesting that context dependence reflects practical rather than fundamental limitations. The consistency of relational geometries across network instances points toward the possibility of universal decoders that approach true $H(I|R)$.

These results establish that neural networks can achieve the unambiguous representations that theoretical accounts of consciousness require. While we make no claims about MNIST networks being conscious, our framework provides a quantitative method for assessing whether a neural substrate satisfies the intentionality constraint, a necessary (though likely not sufficient) condition for phenomenal experience. Future work might apply these methods to biological neural recordings to test whether low representational ambiguity is indeed associated with conscious processing, as the theoretical framework predicts.

\subsection*{Code Availability}
Code and data are available at \url{https://github.com/entropicbloom/consciousness}.

%% file: appendix.tex
\appendix

\section{Representational Alignment and Kernel Similarity}

\subsection{Mutual k-NN Kernel Similarity}

To further validate the connection between our work and the Platonic Representation Hypothesis \citep{huh2024platonic}, we computed the mutual k-NN kernel similarity metric on our MNIST networks. This metric, originally applied to large cross-modal models by Huh et al., measures the alignment of representations across different network instances.

\begin{figure}[htbp]
    \centering
    \includegraphics[width=0.7\textwidth]{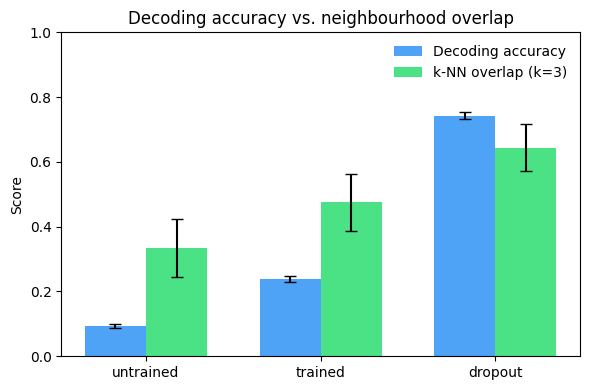}
    \caption{Mutual k-NN kernel similarity versus decoder accuracy across training paradigms. Higher kernel similarity is associated with better decoding performance, suggesting that representational convergence across network instances may reduce ambiguity.}
    \label{fig:knn-similarity}
\end{figure}

We computed the metric by comparing the weight matrices of output neurons across MNIST networks trained with different random seeds. Figure~\ref{fig:knn-similarity} shows the relationship between mutual k-NN kernel similarity and our decoder accuracy. The positive trend suggests that kernel similarity may serve as a proxy for representational ambiguity: networks with higher similarity across instances tend to exhibit more consistent, less ambiguous encoding of class identity.

This relationship suggests that the principles underlying representational convergence in large-scale models also operate in smaller networks. If this pattern generalizes, kernel similarity metrics could provide a computationally efficient proxy for measuring representational ambiguity in larger, multimodal systems where direct decoding approaches may be intractable.